\documentclass[a4paper, amsfonts, amssymb, amsmath, reprint, showkeys, nofootinbib, twoside,aps, prapplied,superscriptaddress]{revtex4-2}

\setcitestyle{numbers,compress,sort&compress}
\usepackage[dvipsnames,table,xcdraw]{xcolor} 
\usepackage[english]{babel}
\usepackage[utf8]{inputenc}
\usepackage[colorinlistoftodos, color=green!40, prependcaption]{todonotes}
\usepackage{amsthm}
\usepackage{mathtools}
\usepackage{physics}
\usepackage{xcolor}
\usepackage{graphicx}
\usepackage[left=23mm,right=13mm,top=35mm,columnsep=15pt]{geometry} 
\usepackage{adjustbox}
\usepackage{placeins}
\usepackage[T1]{fontenc}
\usepackage{lipsum}
\DeclareGraphicsExtensions{.pdf,.png,.jpg,.eps}
\usepackage{csquotes}
\usepackage{hyperref} 

\hypersetup{
	colorlinks=true,
	linkcolor=blue,
	citecolor=blue,
	filecolor=blue,
	urlcolor=blue,
}
\usepackage{color}
\usepackage{ulem}

\newcommand\old{\bgroup\markoverwith{\textcolor{ForestGreen}{\rule[0.5ex]{2pt}{1.6pt}}}\ULon}
\newcommand{\isin}[1]{\sin{ \left( {#1} \right) }}
\newcommand{\icos}[1]{\cos{ \left( {#1} \right) }}

\newcommand{\brakets}[1]
{
	\left\langle  {#1} \right\rangle 
}

\usepackage{natbib}
\begin{document}
	
	\title{Probing atom-surface interactions from tunneling-time measurements via rotation-transport on an atom chip.}

	\author{J-B. Gerent}
	\email{jean-baptiste.gerent@institutoptique.fr}
	\affiliation{Department of Physics and Astronomy, Bates College, Lewiston, ME 04240, USA}%
	\author{R. Veyron}
	\affiliation{%
ICFO - Institut de Ciencies Fotoniques, The Barcelona Institute of Science and Technology, 08860 Castelldefels, Barcelona, Spain%\\This line break forced% with \\
} 
    \author{V. Mancois}
	\affiliation{ Nanyang Quantum Hub, School of Physical and Mathematical Sciences, Nanyang Technological University, 21 Nanyang Link,
 Singapore 637371, Singapore}
 \affiliation{ MajuLab, International Joint Research Unit IRL 3654, CNRS, Université Côte d’Azur,
 Sorbonne Université, National University of Singapore, Nanyang Technological University,
 Singapore}
	\author{R. Huang}
	\affiliation{ LP2N, Laboratoire Photonique, Num\'{e}rique et Nanosciences, Universit\'{e} de Bordeaux-IOGS-CNRS:UMR 5298, rue F. Mitterrand, F-33400 Talence, France}
	
	\author{E. Beraud}
	\affiliation{ LP2N, Laboratoire Photonique, Num\'{e}rique et Nanosciences, Universit\'{e} de Bordeaux-IOGS-CNRS:UMR 5298, rue F. Mitterrand, F-33400 Talence, France}
	
	\author{S. Bernon}
	% 	\author{ $^{1}$, $^{1}$, $^{1}$, and $^{1}$}
	%	\affiliation{$^1$ LP2N}
	\email{simon.bernon@institutoptique.fr}
	\affiliation{ LP2N, Laboratoire Photonique, Num\'{e}rique et Nanosciences, Universit\'{e} de Bordeaux-IOGS-CNRS:UMR 5298, rue F. Mitterrand, F-33400 Talence, France}
	
	\begin{abstract}
	We propose a novel method to measure the interaction between an ultracold gas of neutral atoms and a surface. This solution combines an optical dipole trap reflected by the surface, a magnetic trap formed by current carrying wires embedded below the surface, and a rotation of the surface itself. It allows to adiabatically transport a $^{87}$Rb BEC from few $\mu$m to few hundred nm of the surface.
	{At such distances, atom-surface interaction  strongly affects the trapping potential, causing an increase of the tunneling rate towards the surface.} In this paper, we show that the measurement of the lifetime of the cloud and its comparison to a tunneling model will allow to extract  the Casimir-Polder (CP) force coefficient in the retarded regime ($c_4$).
	{Our model includes noise-induced heating, calibration biases of experimentally controlled parameters and accuracy of the atom lifetime measurement.} Using typical trapping parameters and experimental uncertainties, we numerically estimate the relative uncertainty of $c_4$ to be 10\%.  {This method can be implemented with any atomic species that can be magnetically and optically trapped.}
		
	\end{abstract}
	\date{\today}
	% \keywords{}
	\maketitle

\section{Introduction}

 Ultracold atoms trapped in {optical or magnetic}  potentials are the building blocks of the emerging field of atomtronics \cite{Amico.2021}. Controlling atoms in the vicinity of surfaces enables strong atom-light coupling
 {allowing to explore} quantum regimes where intricate quantum correlations \cite{Gullans2012,Hood2016} and strongly correlated phases \cite{Douglas2015} can be achieved. Reaching an exhaustive control of atomic states and dynamics requires a precise knowledge of atom-surface interactions. Among these, the Van der Waals (VdW) interaction \cite{London1930,Lennard1932} that arises from the interaction between a ground-state atom and a surface has attracted lots of theoretical interest with intriguing predictions for structured surfaces \cite{Buhmann2016}, excited state \cite{Buhmann2016b}, tunability of multilayers surfaces \cite{Antezza2017} or quantum reflection induced levitation \cite{Reynaud2017}.
	
 In this field, theoretical predictions are strongly ahead of experimental realizations which mainly focus on measuring the strength of the interaction and its dependence on temperature \cite{Obrecht2007}. The atom-surface interaction potential is governed by an inverse power law $U_{\rm VdW}=-c_\alpha/z^\alpha$ where $c_\alpha$ is the interaction strength coefficient and $z$ the atom-surface distance. The exponent $\alpha$ has been derived theoretically for various distance range $z$. At short distance, the interaction energy scales as  $\alpha=3$ \cite{Lennard1932} (Lennard-Jones or LJ regime) while it scales as $\alpha=4$ at large distance (Casimir-Polder or CP regime) where retardation effects need to be taken into account \cite{CasimirPolder1948}. The crossover between the two regimes is around $z\approx \lambda/(2\pi)$ where $\lambda$ is the wavelength of the dominant atomic transition involving the ground state. \\
 The experiments, mostly concerned with determining the interaction strength, adopt either a static, dynamic or statistical strategy. Static measurements extract the coefficient from a local measurement of the atomic properties. This was for example realized by measuring the change of center of mass oscillation frequency of a BEC in the vicinity of a surface \cite{Harber2005} or using matter-wave interferometry techniques in a 1D lattice \cite{Balland2024}. While achieving extreme measurement sensitivity, both methods were limited, by the size of the atomic sample, to probe the interaction at large distances ($z>1\,\mu$m, CP regime). On the other hand, dynamic measurements extract the interaction strength from the study of a cold atom cloud scattered by the surface potential. 
 Typical surface geometries that have been probed include planar surfaces using either optical barrier reflection \cite{Landragin1996} or quantum reflection principles \cite{Pasquini2004}, and subwavelength periodically patterned surfaces inducing diffraction of the reflected  \cite{Stehle2011,Bender2014} or transmitted \cite{Sukenik1993,Morley_2021,Garcion2021,Lecoffre2025} matter wave. At short distances, the VdW regime has mainly been probed by direct spectroscopy for hydrogen atoms adsorbed on the surface of liquid helium \cite{Mosk1998} or by frequency selective reflection spectroscopy \cite{Oria_1991,Fichet_2007,Whittaker2014} of alkali atoms in a glass cell. In reflection spectroscopy, atoms from a vapor are probed by the evanescent field of a total internal reflection, thus limiting the probed volume to only a fraction of the optical wavelength.
 
 In this paper, we propose {to probe atom-surface interactions in the CP regime} {by} adiabatically and continuously {transporting} a cloud of ultracold atoms towards a surface. Adiabatic transport typically involves magnetic interaction \cite{Folman.2000,Reichel.2002} or far off-resonance electric dipole interaction driven by an optical field at wavelength $\lambda_0$ \cite{Gustavson.2002,Trisnadi.2022}. As in \cite{Gillen2008}, our solution {combines both} approaches, enabling a continuous control over the atom-surface distances from few {$\lambda_0$} down to $\lambda_0/4$. It is therefore well suited to study atom-surface interactions in the CP regime. It is worth mentioning that our method, which we refer to as \textit{rotation transport}, could be combined at large distance with an adiabatic loading from a magnetic chip trap \cite{Treutlein2004}, and at short distance with  doubly dressed states trapping \cite{Chang.2014,MaximeBellouvet2018}, allowing to reach nanometric atom-surface distances.
We propose here to use this rotation transport to bring a cloud of atoms close to a surface to probe the attractive force of the surface on the atoms. The method relies on measuring the lifetime of the cloud when the tunnel-time toward surface-bonded states becomes predominant \cite{Lin.2004,Petrov.2009,Salem.2010}.\newline 
We present in Fig.\ref{fig:MethodSchematic} a sketch of our method: we plot in (c) 1D cuts along z of the total (plain line) and optical (dashed line) potential at the trap center $\left(x_0,y_0 \right)$. A wave-packet (in green) can, in such potential, tunnel toward other lattice sites (increasing $z$) or toward the surface (at $z=0$). The attractive CP force will mainly decrease the barrier height towards the surface, leading to a shorter {tunneling} time constant $\tau_t$ towards the surface.
The dependency of the {tunneling} time with the surface-dependent CP coefficient and with well-controlled and tunable experimental parameters (surface angle and optical power) is the core of our measurement method.\\

The paper is organized as follow : First, we {introduce} the {rotation} transport method in section \ref{txt:method} with no atom-surface interactions. Then, we present numerical simulation results in section \ref{txt:NumSim} with general trap parameters in \ref{txt:TrapParam} and show in \ref{txt:CPmeas} that CP forces need to be taken into account. By doing so, we highlight how {the rotation transport} method can be used to perform a measurement of surface-atom interaction coefficients. {Furthermore}, we discuss the experimental implementation of our method in section \ref{txt:XPLim}. We derive in \ref{txt:c4precision} the relative precision achievable in typical conditions. {Finally, we} consider some limitations in \ref{txt:adiab} such as three-body losses and transport adiabaticity. 
	
\section{Rotation transport method}\label{txt:method}
	
    \begin{figure}[h!]
		\begin{center}
			\includegraphics[scale=0.4]{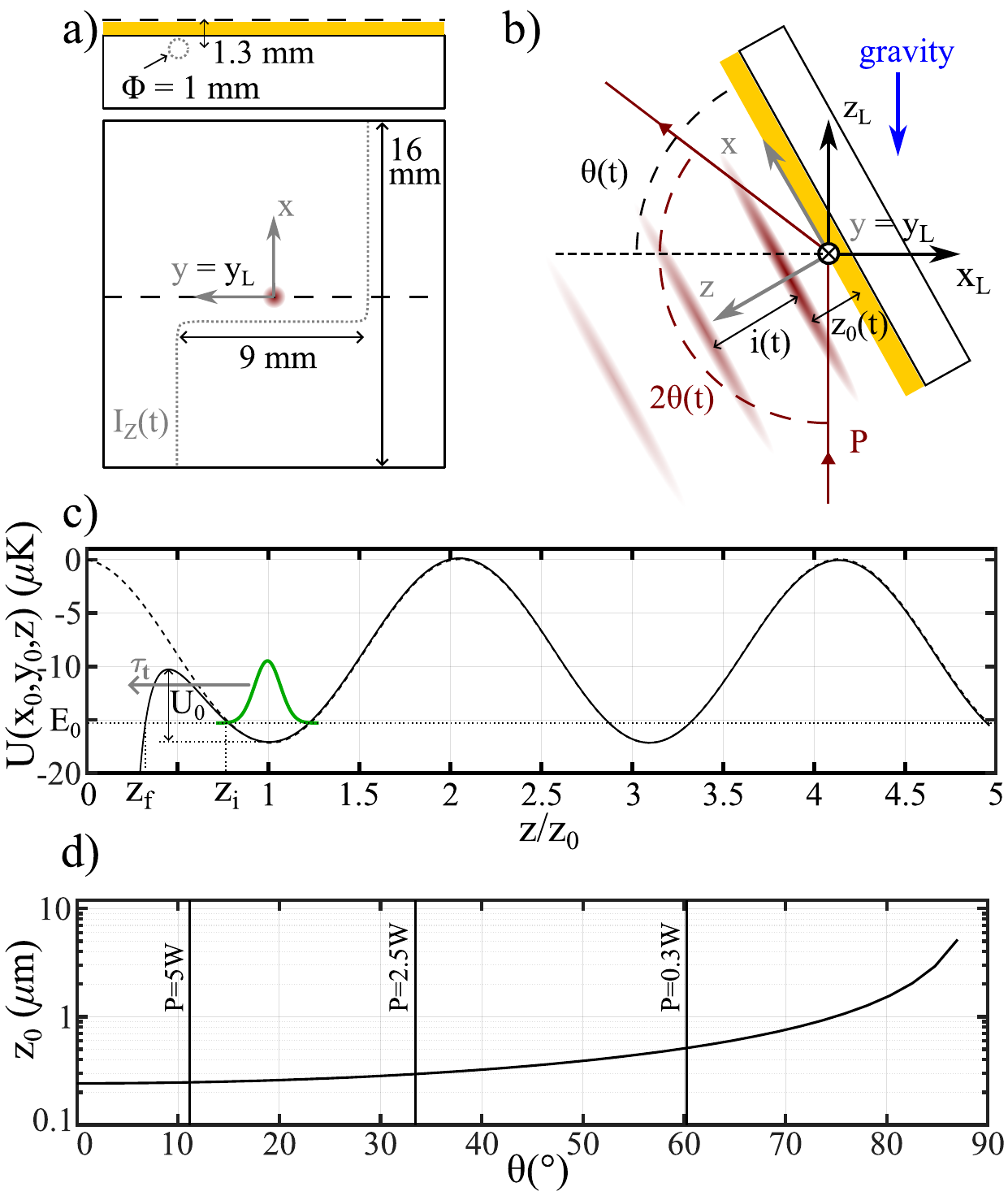}
		\end{center}
		\caption[]{% C:\Users\jgerent\Documents\Bates\LP2N\puce\v_article\fig\fig1v2.svg
		% code is comp_grav2.m for c) and main_fig1d.m for d)  
		a) Drawing of the chip's copper substrate: (top) side view, (bottom) top view. A gold mirror (yellow) is glued on top of the substrate. The rotating axis of the chip (dashed black line, along $y = y_L$) aligns with the top surface of the chip above the mirror. A Z-shaped wire (grey dotted line) carrying a current $I_Z(t)$ is embedded in a groove between the mirror and the substrate. A 1064 nm laser (red fading circle) is reflected by the mirror at the center of the chip (origin of both frames of reference). b) Interference fringes (interfringe $i(t)$) are produced at a distance $z_0(t)$ from the chip by reflecting the laser beam (fixed, propagating along $+z_L$) while rotating the chip, resulting in a time-dependent angle of incidence $\theta(t)$. c) 1D cut of the total (plain line) and optical (dashed line) potential in the perpendicular ($z$) direction at $\theta=65^\circ$. The wave-function (green) can tunnel toward the surface at $z=0$ with a time scale $\tau_t$. d) Semi-log plot of the trap center position $z_0$ as a function of $\theta$ in the absence of CP forces. The vertical lines represent the lowest trapping angle one can reach when including CP forces with $c_4=2.1 \cdot 10^{-55}$ J.m$^4$ for different optical powers $P$. }\label{fig:MethodSchematic}
	\end{figure}

	In this section, we present the {rotation} transport method and introduce how it can be used to precisely measure the $c_4$ coefficient. \newline
	The transport is done by rotating a surface on which a fixed laser beam (propagating {along} $+z_L$), acting as a dipole trap (DT) \cite{Gustavson.2002,Trisnadi.2022} is reflected. We define a frame of reference (FoR) $(x,y,z)$ associated with the surface (see Fig. \ref{fig:MethodSchematic} (a-b), in grey) rotated by the angle $\theta \left( t \right)$ compared to the fixed laboratory (in black) FoR $\left( x_L,y_L,z_L \right)$. The reflecting layer is integrated {on} an atom chip \cite{Hansel.2001} with a Z-shaped current-carrying wire of millimeter size, producing a magnetic gradient, which compensates gravity along one of the weakly trapping transverse direction ($x$).\newline 
	
	We now {concentrate} on the DT generated by the reflected laser. 
	The total intensity is derived in \cite{JBGerent.2024,Sophocles.1999}. We focus here on the simplified case of an incident field polarized along $y$ {(Transverse Electric, TE)}. In this case, the generalized complex Fresnel reflection coefficient is almost constant as a function of the incidence angle ($\left| \rho_{TE}\left(\theta \right) \right| \simeq 1$). The intensity (including the reflection) can be written as :
	\begin{align}
    I(x,y,z) = {} &
    \frac{2 P}{\pi w_0^2} \left(
        e^{-2r_i^2/w_0^2} + e^{-2r_r^2/w_0^2} + \right.  \\
    & \left. 2e^{-\left(r_i^2 + r_r^2 \right)/w_0^2}
        \cos{\left( \frac{4 \pi}{\lambda_0} z \cos{\theta } + \phi_{TE}{\left( \theta \right)} \right)}
    \right)\notag
    \end{align}
	where $r_i^2=\left( x \icos{\theta} - z  \isin{\theta}\right)^2+y^2$ (resp.  $r_r^2=\left( x \icos{\theta} + z  \isin{\theta}\right)^2+y^2$ ) is the radial distance from $\left(x,y,z \right)$ to the propagation axis of the incident (resp. reflected) beam, $\phi_{TE}{\left( \theta \right)}$ is the angle-dependent phase of the Fresnel coefficient, $P$ the optical power of the incident beam and $\lambda_0=1064$ nm the dipole trap wavelength. \newline
	The above intensity presents evenly spaced fringes of period $i\left( t \right)$ in the longitudinal direction ($z$). The transverse profile depends on the incident beam profile which is, {assumed} Gaussian, {of} waist $w_0=150 $ $\mu$m throughout this paper. The first fringe position $r_0=\left( x_0,y_0,z_0 \right)$ is computed numerically. For surfaces with a complex index of refraction, the incident electric field slightly penetrates the surface before being reflected, leading to an angle dependence of the reflected electric field phase ($\phi_{TE}{\left( \theta \right)}$). Albeit this effect, the first fringe position is given with a good approximation in the transverse direction by : \begin{equation} \label{eq:AtChipLightTrapCenterz0}
		z_0{\left( \theta \right) } \simeq \frac{ \lambda_0}{4 \cos{\theta}}
	\end{equation}
	\newline
	As illustrated in Fig. \ref{fig:MethodSchematic} (d), $z_0$ can be reduced by decreasing the incidence angle, thus transporting atoms, previously loaded into one of the fringes, toward the chip. \\	
	The total potential can be written as : 
	
	\begin{align} \label{eq:Vtot}
		U{\left(x,y,z\right)}=&  U_{\rm mag}{\left(x,y,z\right)} + U_{\rm grav}{\left(x,z\right)} \notag \\
		&+ U_{\rm opt}{\left(x,y,z\right)} + U_{\rm CP}{\left(z\right)} \notag \\
		= & \mu_B {\left|B{\left(x,y,z\right)}\right|} \notag \\
		 & + m g {\sqrt{\left(x\sin{\theta}\right)^2+\left(z\cos{\theta}\right)^2}}  \notag \\
		&+ C_{\rm dip} I{\left(x,y,z\right) } - \frac{c_4}{z^4} 
	\end{align} 
	
	where $C_{\rm dip}$ is a parameter proportional to the atomic polarizability and fully defined in \cite[Eqs.10 and 19]{Grimm.1999}. {For} a linearly polarized  beam and {for $^{87}$Rb atoms where we consider} both transitions of the D-line doublet $5^2\text{S}_{1/2} \rightarrow$  $5^2\text{P}_{1/2}, 5^2\text{P}_{3/2}$, it reads : 
	\begin{align}
	    C_{\rm dip}=& \frac{\pi c^2 \Gamma_2}{\omega_2 ^3}\left( \frac{1}{\omega_2-\omega} + \frac{1}{\omega_2+\omega}\right ) \notag \\
	    & + \frac{\pi c^2 \Gamma_1}{2\omega_1 ^3}\left( \frac{1}{\omega_1-\omega} + \frac{1}{\omega_1+\omega}\right )
	\end{align}
	with $c $ the speed of light, $\omega$ the dressing laser angular frequency, $\omega_1$ (resp. $\omega_2$) the frequency between $\ket{5^2\text{S}_{1/2},\rm F=2}$ and $ \ket{5^2\text{P}_{1/2},\rm F=2}$ (resp. $ \ket{5^2\text{P}_{3/2},\rm F=2}$) and $\Gamma_i$ ($i=\left\{1,2\right\}$) the natural linewidth of the transitions. \newline
	At a fixed angle, decreasing the optical power will decrease the {tunneling} time, eventually leading to a case where atoms are not trapped anymore. To indicate the different atom-surface distances that can be reached, this threshold is represented on Fig. \ref{fig:MethodSchematic} (d) as vertical black lines, each annotated with the corresponding optical power. Unless specified, the default value of the CP coefficient is taken as the average of various theoretical values \cite{Bellouvet.2018,Bouscal.2024,Stehle.2011,Friedrich.2002,Stern.2011,Aspect.2003} (considering flat surfaces, both dielectric and metalic) found in the litterature as $c_4=2.1 \cdot 10^{-55}$ J.m$^4$. \\
	By measuring the {atomic} cloud lifetime (when limited by tunnel losses) which depends on the {first fringe} barrier height, one can extract the value of the $c_4$ coefficient. Adjusting the barrier height $\left( U_0\right)$ can be done either by varying the incidence angle or the optical power, which are {two experimentally well controlled parameters.}
	
\section{Numerical model}\label{txt:NumSim}
	We present in this section the results of numerical calculations. In \ref{txt:TrapParam}, we focus on the case without CP forces, presenting the shapes, frequencies, and depth of the trap produced. Then, we show in \ref{txt:CPmeas} how the $c_4$ coefficient can be determined by measuring the lifetime of atoms trapped with the method presented in section \ref{txt:method}.
	
	\subsection{Achievable trap parameters}
	\label{txt:TrapParam}
	\begin{figure}[h]
		\begin{center}
		    \includegraphics[scale=0.4]{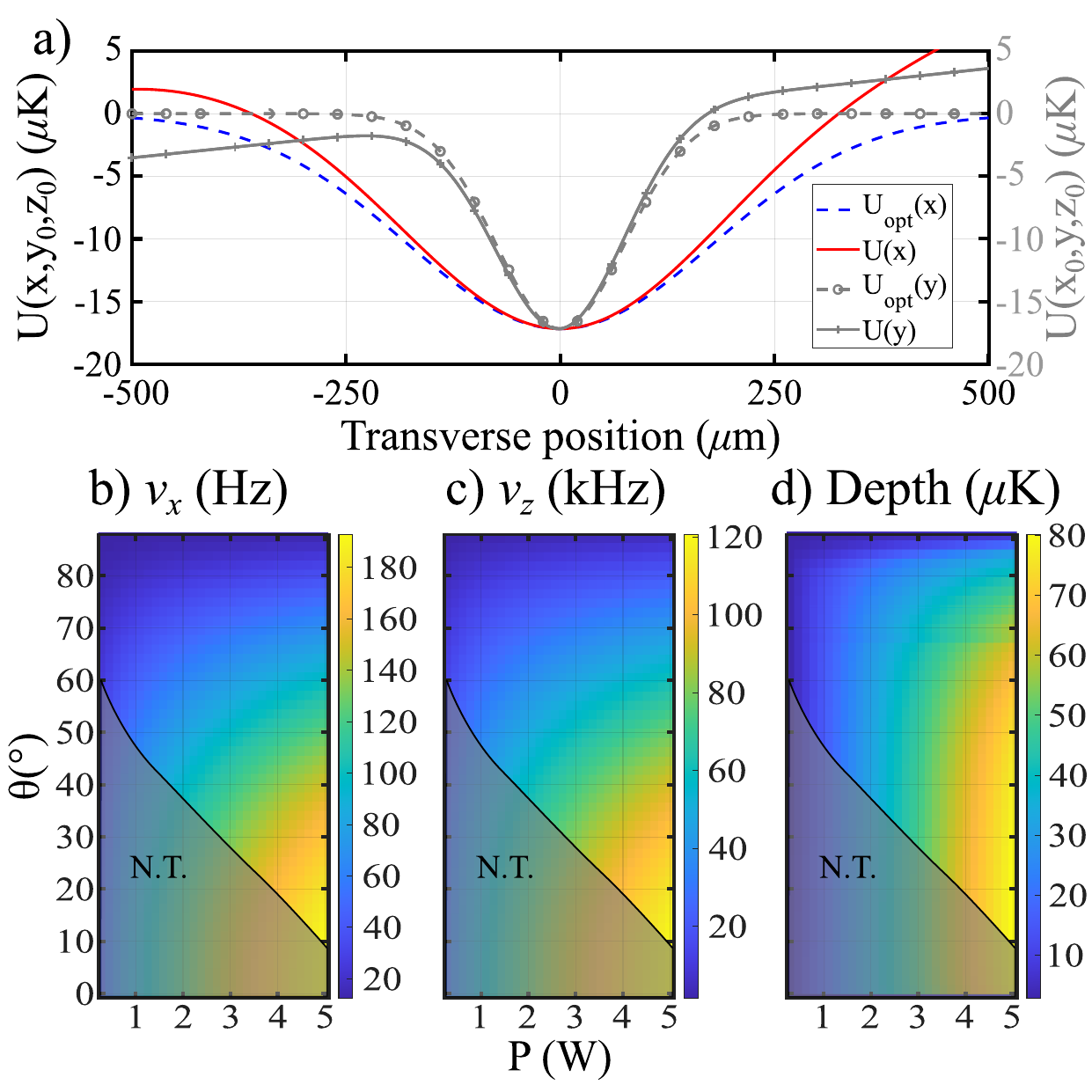}
		\end{center}
		
		\caption{ a) 1D cuts of the potential in the transverse directions. Along $x$ at $(y_0,z_0)$ (left axis, colors, no markers). Along $y$ at $(x_0,z_0)$ (right axis, grey, with markers). 2D maps of trap parameters in the absence of CP forces, as a function of both $P$ and $\theta$ : b) (resp. (c)) Trap frequency in the $x$  (resp. $z$) direction. d) Barrier height ($U_0$). The shaded area  below the black line annotated N.T. (no trap), corresponds to a non trapping configuration when including CP forces with $c_4=2.1 \cdot 10^{-55}$ J.m$^4$. These parameters are achievable nevertheless in the second fringe where the CP force induce very little modification and could be neglected.}\label{fig:TrapParams}
	\end{figure}
	%%%new fig done with main_fig2v4.m for b c d and comp_grav2.m for a 
	The position of the minimum in 3D is numerically computed using the total potential defined in Eq. \ref{eq:Vtot}.
	In the following, we consider a cloud of $^{87}$Rb atoms prepared in $\ket{5^2\text{S}_{1/2},\rm F=2, m_{\rm F}=2}$ for which the s-wave scattering length is $a_0=5$ nm. The complex refractive index \cite[p.249]{Sophocles.1999} used to compute the reflection of the electric field on the surface is $n=\Tilde{n}-\kappa i = 0.25846-6.9654i$ \cite{Yakubovsky.2017}.\newline
    We plot in Fig. \ref{fig:TrapParams} (a) 1D cuts of the potential in the transverse directions ($x$ and $y$) for $\theta= 65 ^\circ$ {in the abscence of CP forces}. Dashed lines represent solely the optical potential ($U_{\rm opt}$). It shows a Gaussian profile elongated in the $x$ direction by the grazing incidence. The current $I_Z(\theta)$ in the Z-wire is computed numerically to maintain the trap center at $x$=0. In practice, it mostly comes down to  compensate the gravity sag along $x$. A $\sin\!{ \left(\theta \right)}$ law is therefore a good approximation for the change of current $I_Z$ with at most $1$\% deviation from the exact calculation. This difference is explained by the fact that the magnetic field generated by a wire is not a pure gradient as required to perfectly compensate gravity but includes higher order terms which cause the asymmetry of $U{\left(x \right)}$ at $|x|>250 ~\mu$m.  To represent a realistic experimental implementation, we have chosen for our simulations a Z-shape trapping geometry which breaks the translational {symmetry} along $y$. Compensating for gravity along $x$, therefore leads to a residual magnetic gradient along  $y$  as shown by the tilt of $U\left( y \right)$ (grey plain line - right axis).\\
    As shown on Fig. \ref{fig:MethodSchematic} (d), reducing the incidence angle ($\theta$) brings the first fringe closer to the surface but also compresses the trap by increasing the trapping frequency in the $x$ (resp. $z$) direction as shown in Fig. \ref{fig:TrapParams} (b) (resp. (c)). To {trustfully} extract a trap curvature from this slightly non-harmonic potential, the trap frequencies are computed with a 1D fit at the bottom of the potential over a length scale being twice the Thomas-Fermi radius along $x$ and $y$ or twice the Quantum Harmonic Oscillator (QHO) length along $z$. 
    The cloud is strongly confined along $z$ such that its size in this direction is reduced and given by the QHO length $a_{\rm QHO}=\sqrt{\hbar /\left(2 \pi \nu_z m \right)}\approx 50 $ nm. {The wavepacket is thus much smaller than the DT wavelength $\lambda_0$, allowing to selectively probe CP force spatially, down to the minimum distance of $\lambda_0/4$ (see Eq. \ref{eq:AtChipLightTrapCenterz0}).} \\
    We plot in Fig. \ref{fig:TrapParams} (d), the trap depth calculated as the minimum of the potential depth in the $x$ and $y$ direction and the barrier height in the $z$ direction. 2D maps of Fig. \ref{fig:TrapParams} are annotated with a black line labeled N.T. (No Trap), below which (grey shaded area) the barrier height $U_0$ is lower than the single particule energy $E_0$ (see Fig. \ref{fig:MethodSchematic}(c)) when including CP force with $c_4=2.1 \cdot 10^{-55}$ J.m$^4$. The tunelling rate increase strongly close to this N.T. region. In the next section, we show that this strong increase of the tunneling can be used to get a precise measurement of the $c_4$ coefficient.
    
	\subsection{Sensitivity on the CP coefficients}\label{txt:CPmeas}
	\begin{figure}[h]
		\begin{center}
			\includegraphics[scale=0.4]{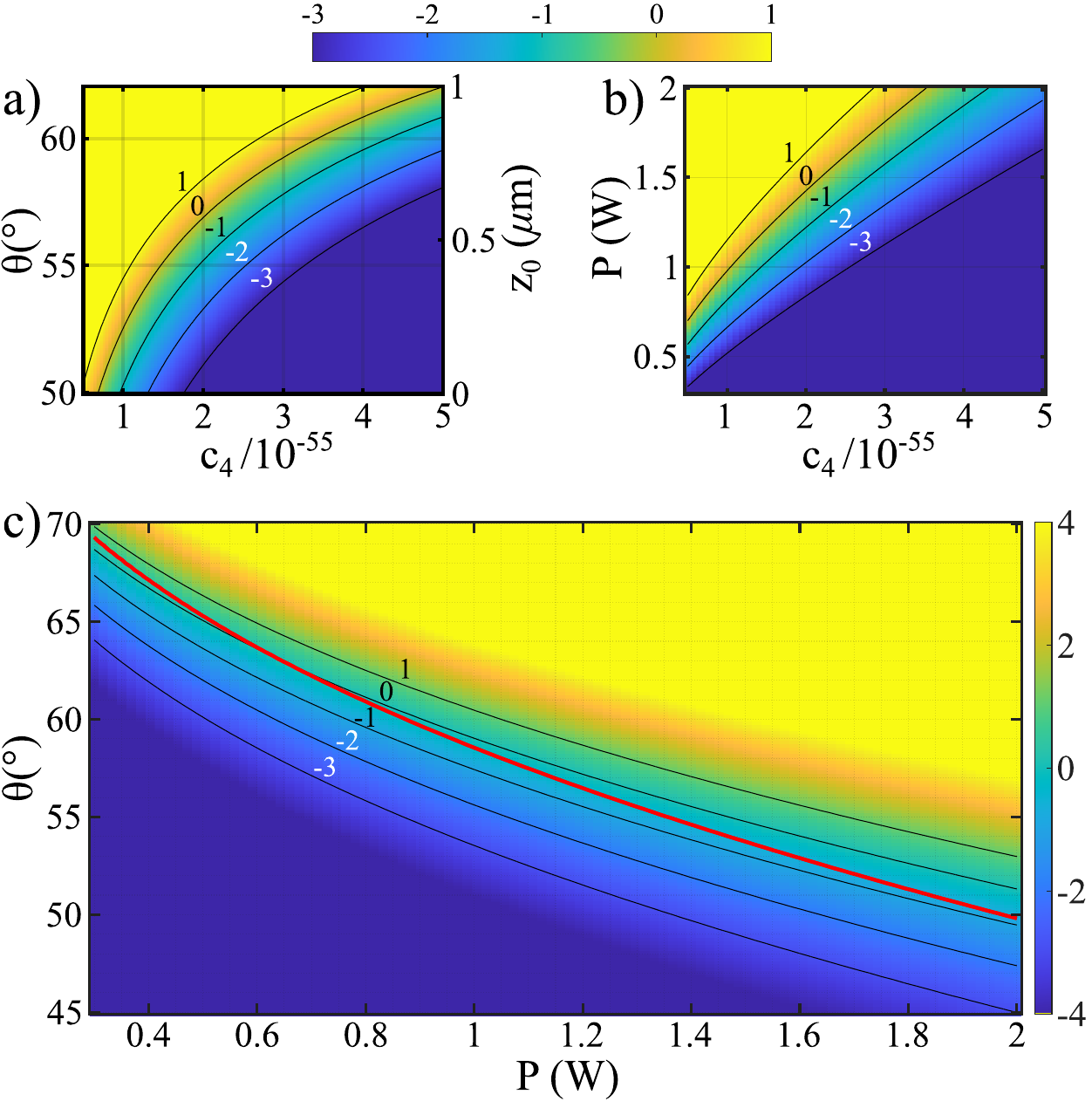}
			%b ) from main_fig3_b_v3.m P =0.3W , 
			% c) from main_fig3_cv2.m do=0.43 microns , 
		\end{center}
		\caption{ 2D maps of the tunnel time, $\text{log}_{10}\left( \tau_t \right)$, limited by the tunneling time toward the surface when including CP forces. Contour lines of annotated values are plotted in black. (a-b) share the same colormap (on top). They are computed at $P=1.2$ W for (a) and $\theta=55 ^\circ$ for (b). The y-axis for (a) (resp. (b)) is $\theta$ (resp. $P$). The right y-axis in (a) shows the equivalent trap center position $z_0$. c) 2D map as a function of $P$ and $\theta$ for $c_4=2.1 \cdot 10^{-55}$ J.m$^4$. The red line corresponds to $\tau_t =\tau_{3b}$ }\label{fig:TauMap}
	\end{figure} 
	In the previous section, we presented general trap parameters as a function of the optical power and incidence angle. {Including the CP force in the calculation of the potential energy (Eq. \ref{eq:Vtot}), modifies the trap profile by reducing the barrier height and shape of the first fringe towards the surface. In the regime where the {atomic cloud} lifetime $\tau$ is limited by tunneling through this barrier, the $c_4$ coefficient can be inferred by measuring $\tau$. This regime is achieved when both the technical source of atom loss (see section \ref{txt:adiab}) and the inelastic three-body losses ($\tau_{3b}$) (see Fig. \ref{fig:NIH}) can be neglected with respect to the tunneling dynamics.} The {tunneling} time is derived in the Wentzel-Kramers-Brillouin (WKB) approximation \cite{Griffiths.2018}. The probability for an atom to tunnel is given by : \begin{equation}
	    P_t=\frac{4}{e^{2\gamma}\left( 1-e^{-2\gamma}\right)^2}
	\end{equation}
	with $\gamma=\frac{1}{\hbar} \int_{z_i}^{z_f}\sqrt{2 m \left| U\left(z'\right) - E_0\right|} dz'$ where $E_0$ is the single particule energy, $z_i$ and $z_f$ are the position at which $U\left ( z \right ) = E_0 $ (see Fig. \ref{fig:MethodSchematic}(c)). \newline
	The tunnel time is then given by : 
	\begin{equation}\label{eq:taut}
	    \tau_t=\frac{1}{P_t \nu_z}
	\end{equation}
	We calculated $E_0$ as the energy of the ground state of a 1D QHO along $z$. Beyond the harmonic approximation, numerically solving the stationary Schrödinger equation using the numerical potential only induces a relative correction of $\Delta E_0/E_0\approx 10^{-2}$ for $\tau_t > 100$ ms. The WKB approximation is then valid for tunnel time above 10 ms with a relative correction $\Delta E_0/E_0\approx 5 \cdot 10^{-2}$.
	
	We plot in Fig \ref{fig:TauMap}, 2D maps of $\log_{10}{\left( \tau_t\right)}$ as a function of $c_4$ and the incidence angle (resp. the optical power) in (a) (resp. (b)). We scan in (c) both the optical power and the angle for a fixed value of $c_4=2.1 \cdot 10^{-55}$ J.m$^4$. The red line corresponds to $\tau_t =\tau_{3b}$ as defined in section \ref{txt:adiab} and calculated here for $N=2\cdot 10^4$ $^{87}$Rb atoms in a pure BEC state. {The {tunneling} limited regime is therefore reached for} $\tau_t \ll \tau_{3b}$. We see on this figure that $\tau_t$ depends both on $P$, $\theta$ and $c_4$, and is strictly monotonous with $c_4$ for fixed values of $P$ and $\theta$. In theory, if $P$ and $\theta$ are {absolutely} known, a single measurement of the trap lifetime would determine $c_4$ {with no ambiguity}. In this case, the strong sensitivity of $\tau_t$ with $c_4$ will help determine $c_4$ precisely. However, uncertainties on ($P$,$\theta$,$\tau$) will affect {the final uncertainty on the exctracted value of} $c_4$.\\
	Formally, we introduce a numerical function which truthfully represents $c_4$ as $c_4^{\rm n}= c_4^{\rm n}\left(\tau,\theta, P \right)$. 
	The $c_4^{\rm exp}$ coefficient measured experimentally can be written as 
	$c_4^{\rm exp}= c_4^{\rm n} +  \delta c_4$, where $\delta c_4$ is the uncertainty on the measurement of the actual $c_4^{\rm n}$ coefficient.\\
	Within this formalism, the sensitivity of the $c_4^{\rm n}$ coefficient with respect to $\theta$ (or one of the other parameter) can be numerically derived at constant $\tau_t$ and $P$ using a variational approach.
	
	Since the slope of vertical cuts of Fig. (a-b) (at constant $c_4^{\rm n}$) depends on the value of $c_4^{\rm n}$, measuring the lifetime while scanning either $\theta$ or $P$ allows to go beyond this single shot method and circumvent systematic errors on the experimental parameters.\newline
	The independence of $c_4$ with respect to $z$ can be verified by repeated measurement at different $z_0$ by scanning $\theta$ (and possibly adjusting $P$ to keep the lifetime in the correct range). 
	Using these curves, the sensitivity can be extracted with the variational method mentioned above. In section \ref{txt:c4precision}, we give the expected sensibility for a realistic experimental implementation.
	
\section{Discussion on experimental implementation}\label{txt:XPLim}
	In this section, we discuss some specific care that should be taken in the experimental implementation of our method. \newline
	We first evaluate the uncertainty on the CP coefficient using typical experimental stability of the different parameters involved such as the beam waist, the incidence angle and the optical power. Then, we discuss two caveats regarding three body losses and transport adiabaticity.
	
	\subsection{Precision on the CP coefficient}\label{txt:c4precision}
	In this section, we set an upper bound on our $c_4$ measurement precision. For this, we measure experimental uncertainties of the parameters ($\theta$, $P$, $w_0$) included in our model (see Eq.\ref{eq:Vtot}), and consider typically achieved experimental measurement precision of the atomic cloud lifetime to extrapolate a typical uncertainty on $c_4$ for the one shot measurement.
	
	Assuming small and uncorrelated fluctuations of experimental quantities $\left(\tau \right)$ and variables $\left(P, \theta \right)$, the relative standard deviation is given by:
	\begin{align} \label{eq:sensi}
	    \frac{\delta c_4}{c_4^{\rm n}}=& \sqrt{\epsilon_{\theta}^2+\epsilon_{P}^2+\epsilon_{\tau}^2}\\ =& \sqrt{\left( \frac{\partial c_4^{\rm n}}{\partial \theta} \frac{\delta \theta}{c_4^{\rm n}} \right)^2+\left( \frac{\partial c_4^{\rm n}}{\partial P} \frac{\delta P}{c_4^{\rm n}} \right)^2+\left( \frac{\partial c_4^{\rm n}}{ \partial \tau} \frac{\delta \tau}{c_4^{\rm n}} \right)^2 \notag}
	\end{align}
	Every experimental parameter can be decomposed as its true value ($\theta_0$), a systematic error ($\theta_{\rm syst}$) and {a statistical error} ($\delta\theta$): $\theta=\theta_0+\theta_{\rm syst} +\delta\theta$.\\
	To experimentally rotate the surface, we consider using a continuous-wave motor (details in \ref{txt:adiab}). The optically encoded angular position and the closed-loop control system results in an uncertainty of $\delta\theta= 273$ $\mu$rad. By retroreflecting the beam, absolute calibration of the angle of the beam with respect to the surface can be easily reduced to less than $\theta_{\rm syst}= 150 $ $\mu$rad. The maximum error we consider here is then $\theta_{\rm syst}+\delta\theta= 423$ $\mu$rad.\newline
	The tunneling time depends on the barrier height, which is experimentally set by the optical intensity $I \propto P / w_0^2$. To first order, fluctuations of the intensity can be treated as effective fluctuations of the optical power. \newline
	Uncertainty on the optical power are experimentally limited by the absolute calibration of optical detectors. A standard calibration uncertainty given by the manufacturer for \hyperlink{https://www.thorlabs.com/thorproduct.cfm?partnumber=S120C}{ThorlabsS120C} is $\frac{P_{\rm syst} +\delta P}{P_0}= 6 \cdot 10^{-2}$. \newline
	Small fluctuations of the waist are experimentally equivalent to power fluctuations. Using forced oscillations \cite{Pitaevskii.1996, JBGerent.2024}, the in-situ dipole trap waist can be measured with a relative precision of ${\delta w_0}/{w_0}=4 \cdot 10^{-3}$. This is equivalent to a power fluctuation of ${\delta \Tilde{P}}/{P}=2 {\delta \omega_0}/{\omega_0} = 8 \cdot 10^{-3}$. The square of which is negligible compared to $\left({\delta{P}}/{P}\right)^2$. The total uncertainty is then $\epsilon_P=6 \cdot 10^{-2} $. \\

	We present here {upper} bounds on the different sensitivities we studied numerically, subject to optimization depending on the expected value of $c_4$. \newline
	{We consider an experimental implementation centered around the parameters : $P=0.5 $ W, $w_0=150\ \mu$m, $\theta= 55 ^\circ$.}   For $\theta \in \left[52,75 \right] ^\circ$ and $\tau_t \in \left[10^{-3},10^{-2} \right] $ s, the relative error contribution from angle uncertainty is $\epsilon_\theta < 20 \times  423 \cdot 10^{-6} = 10^{-3}$. At $\theta= 55 ^\circ$, the contribution from the optical power ($P\in \left[0.3,2 \right] $ W) is $\epsilon_P  < 10^{-1} $. \newline 
	We have observed numerically that $\frac{\partial c_4^n}{\partial \tau} \frac{\tau}{c_4} \approx 0.1$ is mostly independant of $\tau$. Thus, the contribution from the lifetime is 10 time smaller than the relative error on the lifetime ($\epsilon _\tau =\frac{\partial c_4^n}{\partial \tau} \frac{\tau}{c_4} \frac{\delta \tau}{\tau}  < 0.1 \frac{\delta \tau}{\tau} $) and can be statistically reduced. Even for $\frac{\delta \tau}{\tau}=0.1$, the total uncertainty would still mainly comes from $\epsilon_P$.  \newline 
	According to our error budget, the main source of error comes from the uncertainty on the optical power. 
	A calibration from \hyperlink{https://shop.nist.gov/ccrz__ProductDetails?sku=42110C&cclcl=en_US}{NIST}, yields an uncertainty at $1064$ nm of $\delta P / P < 2 \cdot 10^{-3}$. This reduce relative error down to $\frac{\delta c_4}{c_4}= 1.7 \cdot 10^{-2}$ where all uncertainties have comparable contributions.\newline
	We emphasize that measuring the lifetime while scanning one of the parameters allows to be insensitive to systematic errors and could further reduce the uncertainty.
	%See c4_sensitivityv2.m section \epsilon_\tau (l.111)%
	
	\subsection{Additional loss mechanism}\label{txt:adiab}
	In this section, we discuss how two additional loss terms (inelastic collision and adiabatic transport) affect the lifetime measurement and therefore our estimate of $c_4$. First, we consider inelastic collisions and then, limitations on the transport adiabaticity (ability to transport an initial cloud without excitations or significant atom losses). This is critical since the {tunneling} time depends on the trapped cloud energy ($E_0$). \newline
	
\begin{figure}[h!]
		\begin{center}
			\includegraphics[scale=0.4]{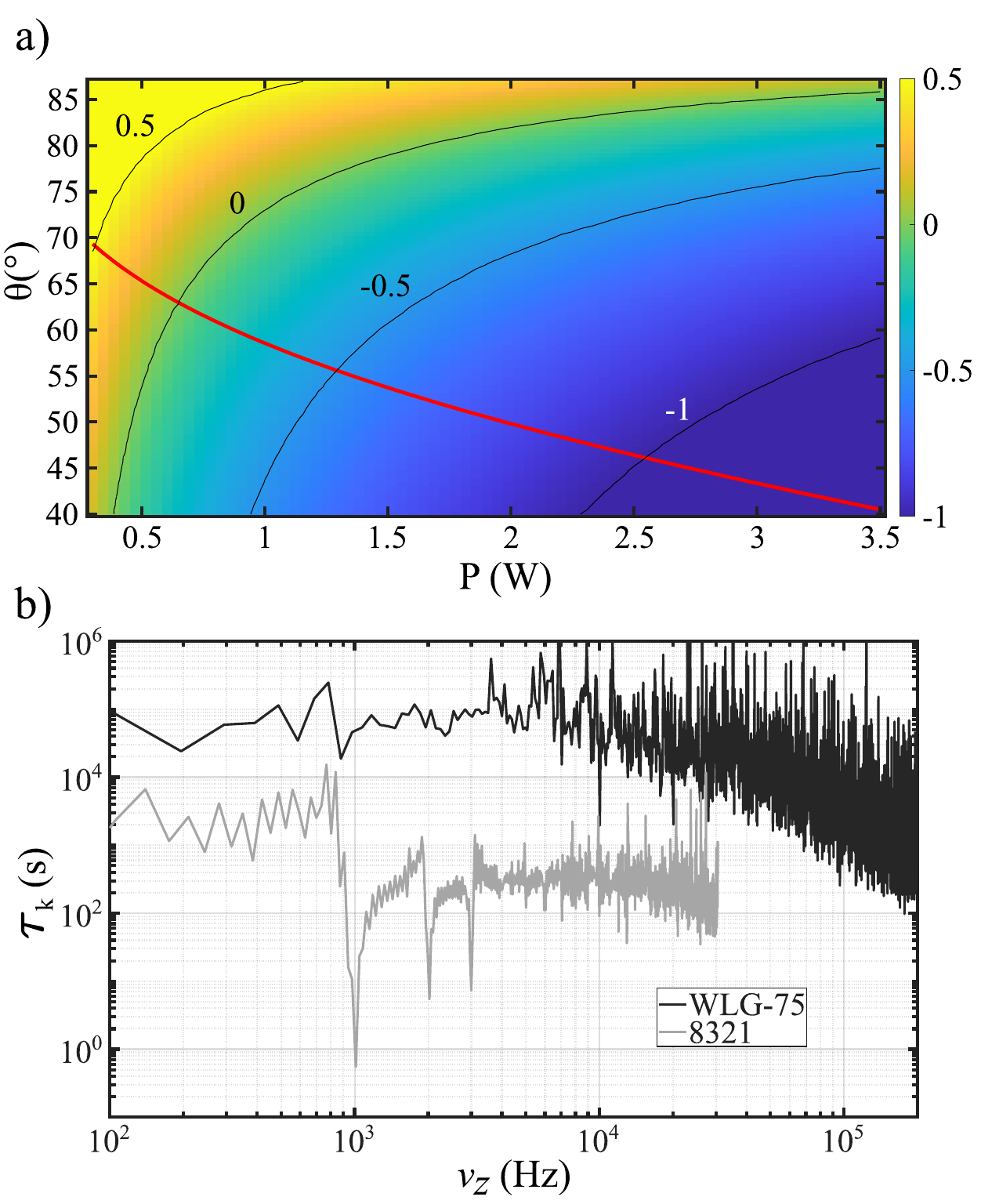}
			%Data from : C:\Users\jbeu\Documents\iogs\thesis_manuscript_fresh\matlab\AtChip\Newport\19_05_20\balayage_phd.m exported in mini window on mian screen
		\end{center}
		\caption[]{a) 2D map of the three-body recombination time ($log_{10}\left( \tau_{\rm 3b} \right)$) as a function of $P$ and $\theta$. Contour lines of annotated values are plotted in black. The red line corresponds to $\tau_{\rm 3b}=\tau_{\rm t}$.  b) Fluctuation-limited lifetime ($\tau_k$) as a function of the trapping frequency ($\nu_z$). The black curve is for a continuous rotation motor (Teckceleo). The grey curve is for a stepper motor (New Focus), driven at $2$ kHz. }\label{fig:NIH}
	\end{figure}

	\textbf{Three-body losses}: \\
	Experimentally, the inelastic three-body collision rate increases during the rotation because the change in the trap frequencies (see Fig. \ref{fig:TrapParams} (b-c)) leads to the increase of the atomic density. We consider a BEC with $2 \cdot 10^{4}$ atoms and plot a 2D map in Fig. \ref{fig:NIH} (a) of the associated time $\tau_{3b}=1/\left(5.8\cdot 10^{-30} \times n_0^2\right)$ (s) \cite{Burt.1997} where $n_0$ is the atomic density in at.cm$^{-3}$.\\
	To avoid this limitation, experiments should be conducted in a regime where $\tau_t \ll \tau_{3b}$. \newline
	
	\textbf{Adiabatic transport}: \\ 
	Our model considers that every particle of the cloud is in the ground state of an ideal QHO. Under this asumption, we can exclude anharmonic couplings, that would mostly impact excited states by hybridizing them and would require to account for dynamical tunneling \cite{Davis1981}. Our calculation therefore {assumes} a high {ground state} preparation fidelity followed by an adiabatic transport.\\
	
	Using the formalism developed in \cite{Savard.1997,Gehm.1998}, we evaluate now the heating of {the atomic} cloud due to fluctuations of the different experimental parameters such as laser stability (polarization, power, pointing) and surface position (both angle and absolute position).\\
	We consider here a 1D problem where the different noises on each experimental parameters are uncorrelated and the time dependence of the QHO parameters ($\epsilon_k,\epsilon_z$) are treated independently. For noises affecting both the trap center position and trap frequencies, each heating term is calculated independently and compared. The independence hypothesis is valid when the corresponding heating times are separated by several orders of magnitude, which is the case for the data presented  below. {Out of the experimental parameters, we present below only the contributions of the most critical parameter: the angular positioning of the chip, driven by the motor.} The 1D hypothesis is justified in this case since {only} $\nu_z$ is {significantly} affected {and because it induces heating over the shortest time scale}. The total heating rate can be generalized {to} higher dimensions as the average of the ones obtained by treating each spatial dimensions {separately} \cite[3.1.4.1.]{JBGerent.2024}. \\
	Following \cite{Savard.1997,Gehm.1998}, we consider the Hamiltonian of the form: 
	\begin{equation}
		H \left( t \right)  = \dfrac{1}{2}m \omega_z^2\left( 1+\epsilon_k\left( t \right)\right) \left( z - \epsilon_z \left( t \right) \right) ^2
	\end{equation}
	where $\epsilon_k$ (resp. $\epsilon_z$) is the fractional fluctuation in the spring constant (resp. trap center position) studied using perturbation theory. 
	
	Fluctuations on the spring constant lead to an exponential increase of the average energy with a {constant rate} $\Gamma_k$ :
	\begin{equation} \label{eq:AtChipGammak}
		\Gamma _k =\frac{1}{\tau_k}= \pi ^2 \nu _z ^2 S_k\left( 2 \nu _z \right) 
	\end{equation}
	where $\nu _z $ is the trap frequency and $S_k$ is the one-sided power spectrum of the fractional fluctuation in the spring constant normalized so that : 
	\begin{equation} 
		\int_0^\infty S _k\left(\nu _z \right) = \epsilon_{0,k}^2
	\end{equation}
	with $\epsilon_{0,k}$ the root-mean-square fractional fluctuation in the spring constant.
	
	Fluctuations on the trap center position lead to a heating rate independent of the trap energy. We define an energy doubling time $\tau_z$, as the time needed to double the initial average energy :
	\begin{equation}\label{eq:AtChipTx}
		\tau_z=\frac{\brakets{E_z{\left( t=0\right) }}}{4 m\pi^4 \nu_z^4 S_z{\left( \nu_z\right) }}
	\end{equation}
	where $\brakets{E_z{\left( t=0\right) }}$ is the initial average energy, $m$ is the particle mass and $S_z$ is the one-sided power spectrum of the {fractional} fluctuation in the trap center position.\\
	Fluctuations in the incidence angle lead to fluctuations {in} both the trap frequency and position:
	
	\begin{equation}\label{eq:AtChipEps}
\left\{
\begin{aligned}
\epsilon_k\left(\theta \right) &= 2\delta \theta\, \tan\!{\left(\theta \right)}\\
    \epsilon_z\left(\theta \right) &= \delta \theta \, z_0\! \left(\theta \right) \tan\!{\left(\theta \right)} 
\end{aligned}
\right.
\end{equation} 
    with $\delta \theta$ the angle fluctuations and $z_0\! \left(\theta \right)$ the distance to the surface defined in Eq.\ref{eq:AtChipLightTrapCenterz0}. \\ 
	Because the energy increase from $\epsilon_k$ is {the dominating one} at a constant time, we show only the results of the study of $\epsilon_k$.
	Extra care {is} required during the initial stage of the transport due to the dependence in $\tan\!{\left(\theta \right)}$, which diverges for grazing incidence.\\ 
	
	We considered two {distinct types of} motors, one piezo-driven stepper motor (\hyperlink{https://www.newport.com/p/8321}{New Focus picomotor 8321}) and a continuous rotation motor (\hyperlink{https://www.tekceleo.com/technologies/piezoelectric-motor/}{Tekceleo WLG-75}).\\
	We plot in Fig. \ref{fig:NIH} (b) the two heating times for both these motors as a function of the trap frequency, computed using  Eqs. \ref{eq:AtChipGammak} and \ref{eq:AtChipEps} . The stepper motor (grey curve) is globally more noisy and presents problematic harmonics at half of the drive frequency (1kHz), that would constraints the experiment.  The noise spectrum of the WLG-75 sets an upper bound on the energy increase over the 100 ms long transport of a cloud with a maximum trap frequency of 20 kHz ($E_0/h$) as : $\Delta E/h=E_0/h \left( e^{\Gamma t} -1 \right) < 10^{-4} \cdot 20$ kHz. This contribution is  below the theoretical heating due to the scattering of photons from the dipole trap laser field \cite{Grimm.1999} $\Delta E_{\rm DT}/h = 3 \cdot 10^{-3} \cdot 2$ kHz. Both lead to a negligeable energy increase: $\Delta E / E_0 = 3 \cdot 10 ^{-4} \ll 1$ and the limitation of the lifetime by noise-induced heating and photon scattering can therefore be disregarded.
	
	\section{Conclusion and Outlook}
	In this paper, we showed that using a reflected optical dipole trap on a rotating atom chip allows to precisely measure the Casimir-Polder (CP) coefficient of a reflecting surface. This is achieved by measuring atom losses induced by the distance-dependent tunneling through the potential barrier towards the surface. By rotating {the} surface, it is possible to probe the interaction at various distances and could be used to verify the $1/z^\alpha$ scaling law ($c_4$ being independent of $z$), and potentially probe the transition from the CP (retarded) to the Lennard-Jones regime. From our preliminary  evaluation, a relative uncertainty of $\delta c_4 / c_4 =0.1$ can be {achieved} and could be reduced to a few percent with efforts on the calibration of the optical power and by optimizing $\left(P,\theta\right)$ to minimize the sensitivities introduced in Eq. \ref{eq:sensi} for a targeted $c_4$.
	
	Other measurement techniques of the atom-surface interaction can also be implemented using our scheme, {for example by} measuring the modification of the trap frequency due to the CP force  \cite{Harber2005}. This frequency could be extracted by measuring the oscillation frequency of the center of mass of the cloud or by measuring the lifetime after exciting a parametric oscillation, {or even} transfering the cloud to a higher vibrational state with a significantly shorter tunnel-time. In our trapping potential, we estimate a normalized oscillation frequency shift on the order of $10^{-2}$, well above the $10^{-4}$ reported in \cite{Harber2005}, where the effect was already within experimental measurement precision.
	
	The determination of the $c_4$ coefficient depends on the precise knowledge of the total trapping potential. Out of the four effects involved in Eq. \ref{eq:Vtot}, the dipole force is the main source of error. Indeed, gravity and the magnetic potential can be precisely simulated for a correctly defined geometry of the trapping wires and coils structures. The {exact} magnetic field can also be measured {via} microwave spectroscopy. To go beyond the numerical model we implemented, the light shift due to the laser field could be computed by diagonalizing the Hamiltonian $H=H_{\rm Stark} + H_{\rm hf}$ composed of the sum of the AC Stark Hamiltonian and the hyperfine shift Hamiltonian. Such diagonalization leads to  relative corrections on $U_{\rm opt}$ that are typically below $10^{-4}$.
	
	The atom-surface interaction primarily depends on the atom-surface distance, but also on the surface and environment properties. At distances on the order of the thermal wavelength (typically a few \textmu{m}), thermal fluctuations from the surface and its environment contribute to the Casimir-Polder force \cite{Antezza2005,Obrecht2007}. Another important systematic effect arises from free electrons and adsorbate-induced electric fields, that are responsible for spatially inhomogeneous energy shifts \cite{Hattermann.2012,Sedlacek.2016}. Since our rotation transport method will be sensitive to these systematics, it could be used to measure such contributions to the total potential. Also, compensation techniques can be envisionned, such as double-dressing \cite{Bellouvet.2018} or photodissorption methods \cite{Hattermann.2012} to prevent surface adsorption. At distances of tens of \textmu{m}, the bias electrostatic force can be measured and accounted for in the model for precise measurement of the Casimir-Polder force in the \textmu{m} range \cite{Balland2024}. For distances below the micrometer, the Casimir-Polder force will dominate. \\
	The rotation transport method introduced in this paper is therefore suitable to precisely and locally measure atom-surface interaction over a large range of distances.
	\vspace{1cm}
    
	\bibliography{main_arxiv.bib}
	
\end{document}